\documentclass[12pt]{article}
\pdfoutput=1
 \def\be{\begin{equation}}
 \def\ee{\end{equation}}
 \def\bea{\begin{eqnarray}}
 \def\eea{\end{eqnarray}}
 \usepackage{graphicx}
\usepackage{amsmath,amssymb}
\usepackage{hyperref}
\usepackage{color}

 \catcode`\@=11
 \def\lsim{\mathrel{\mathpalette\@versim<}}
 \def\gsim{\mathrel{\mathpalette\@versim>}}
 \def\@versim#1#2{\vcenter{\offinterlineskip
 \ialign{$\m@th#1\hfil##\hfil$\crcr#2\crcr\sim\crcr } }}
 \catcode`\@=12

 \catcode`@=12
\begin{document}
 \thispagestyle{empty}
 \begin{flushright}
 UCRHEP-T579\\
 June 2017\
 \end{flushright}
 \vspace{0.5in}
 \begin{center}
 {\LARGE \bf Anomalous Leptonic U(1) Symmetry:\\ 
 Syndetic Origin of the QCD Axion,\\ 
 Weak-Scale Dark Matter,\\ and Radiative Neutrino Mass\\}
 \vspace{0.5in}
 {\bf Ernest Ma$^1$, Diego Restrepo$^2$, and \'Oscar Zapata$^2$\\}
 \vspace{0.2in}
 {\sl $^1$ Department of Physics and Astronomy,\\ 
 University of California, Riverside, California 92521, USA\\}
\vspace{0.1in}
{\sl $^2$ Instituto de F\'isica, Universidad de Antioquia,\\ 
Calle 70 No. 52-21, Apartado A\'ereo 1226, 
Medell\'in, Colombia\\} 
\end{center}
 \vspace{0.5in}

\begin{abstract}\
The well-known leptonic U(1) symmetry of the standard model of quarks and 
leptons is extended to include a number of new fermions and scalars.  
The resulting theory has an invisible QCD axion (thereby solving the 
strong CP problem), a candidate for weak-scale dark matter, as well as 
radiative neutrino masses.  A possible key connection is a color-triplet 
scalar, which may be produced and detected at the Large Hadron Collider.
\end{abstract}

 \newpage
 \baselineskip 24pt
\noindent \underline{\it Introduction}~:~
In the standard model (SM) of quarks and leptons, there are four automatic 
global symmetries: baryon number $B=1/3$ for each quark, lepton number 
$L_e=1$ for the electron and its neutrino $\nu_e$, $L_\mu=1$ for $\mu$ and 
$\nu_\mu$, and $L_\tau=1$ for $\tau$ and $\nu_\tau$.  As such, all neutrinos 
are massless. Given that we now know that neutrinos are massive and mix 
among themselves, the SM must be extended and $L_{e,\mu,\tau}$ must be replaced 
with $L = L_e + L_\mu + L_\tau$.  Hence $L$ is still a valid global $U(1)$ 
symmetry if neutrinos are strictly Dirac fermions, but often than not, 
they are assumed to be Majorana fermions so that $L$ is broken to 
$(-1)^L$, i.e. lepton parity.

Theoretical mechanisms for obtaining Majorana neutrino masses are 
many~\cite{m98}, but there is no experimental evidence for any one of them.  
Then there is the dark matter (DM) of the Universe.  The SM has no explanation 
for it, but the intriguing idea exists that it may be connected to the 
neutrino's mass generating mechanism.  In 2006, a simple one-loop radiative 
mechanism was proposed~\cite{m06} with dark matter in the loop, called 
``scotogenic'' from the Greek ``scotos'' meaning darkness.  In 2015, it was 
shown~\cite{m15} that the dark parity of this model, as well as many 
others, is derivable from lepton parity.  This demonstrates how the 
leptonic $U(1)_L$ symmetry may be extended to include particles beyond 
those of the SM.

In 2013, it was shown~\cite{dmt14} that the well-known spontaneously 
broken anomalous Peccei-Quinn $U(1)_{PQ}$ symmetry~\cite{pq77}, which 
solves the strong CP problem~\cite{kc10} and creates the QCD 
axion~\cite{we78,wi78}, has a residual $Z_2$ symmetry which may in fact 
be dark parity.  In this paper, we combine all these ideas to show 
that, with the proper choice of fermions and scalars beyond the SM, we
can have $U(1)_{PQ} = U(1)_{L}$ with the residual dark parity~\cite{m15} 
given by $(-1)^{3B+L+2j}$, i.e. the well-known $R$ parity of supersymmetry, 
but not in a supersymmetric context.

To implement this important new insight, i.e. $U(1)_{PQ}=U(1)_L$, in a 
specific model, we choose the singlet-doublet-fermion dark-matter 
scenario with three additional scalars to obtain scotogenic neutrino 
masses~\cite{fmp14,rrszt15}.  This framework is however also adaptable for 
radiative quark and lepton masses~\cite{m14,m15-1}.

\noindent \underline{\it Particles Beyond the SM}~:~
The new particles of our model are assigned under $U(1)_L$ as shown in Table 1.
\begin{table}[htb]
\caption{Particle assignments under $PQ=L$.}
\begin{center}
\begin{tabular}{|c|c|c|c|c|c|c|}
\hline
Particle & $SU(3)_C$ & $SU(2)_L$ & $U(1)_Y$ & $PQ=L$ & $B$ & $R$ \\
\hline
$D_L$ & 3 & 1 & $-1/3$ & $1$ & 1/3 & $-$ \\
$D_R$ & $3$ & 1 & $-1/3$ & $-1$ & 1/3 & $-$ \\
\hline
$N_L$ & $1$ & 1 & $0$ & $0$ & 0 & $-$ \\
$(E^0,E^-)_{L,R}$ & 1 & 2 & $-1/2$ & $0$ & 0 & $-$ \\
\hline
$\zeta$ & 3 & 1 & $2/3$ & $0$ & 1/3 & $-$ \\
\hline
$\chi_{1,2,3}$ & 1 & 1 & 0 & 1 & 0 & $-$ \\
\hline
$\sigma$ & 1 & 1 & 0 & 2 & 0 & + \\ 
\hline
\end{tabular}
\end{center}
\end{table}
Note that the only new fermion which transforms under $U(1)_L$ is $D$.  
As this anomalous $U(1)_L$ is broken by 
$\langle \sigma \rangle$, the QCD axion appears, together with the 
residual symmetry $R = (-1)^{3B+L+2j}$, which is even for SM particles as well 
as $\sigma$, but odd for all the other new particles.  The axion is thus 
of the KSVZ type~\cite{k79,svz80} and the domain wall number is 1, so it 
is cosmologically safe~\cite{s82}.

The axion decay constant $F_A$, i.e.  
$\langle \sigma \rangle$, is known to be large~\cite{rs88}: 
$F_A > 4 \times 10^8$ GeV.  Hence the singlet $D$ quark is expected to 
be heavy, unless the Yukawa coupling for $\sigma \bar{D}_L D_R$ is very 
small.  If this is indeed the case, then $D$ may be produced in pairs at 
the Large Hadron Collider (LHC), and the $\bar{D}_L d_R \chi$ 
term~\cite{dmt14} would 
allow it to be discovered.  Alternatively, if a dark scalar doublet 
$(\eta^+ \eta^0)$ exists as in the original scotogenic model~\cite{m06}, 
then the $\bar{D}_R (u_L \eta^- + d_L \bar{\eta}^0)$ term works 
as well~\cite{acdlnq16}.  On the other hand, if $D$ is very heavy 
(of order $F_A$) as expected, then it is impossible for it to be 
produced at the LHC.  In this study, we will consider instead the dark 
scalar quark $\zeta$ with charge $2/3$.  Its mass is not constrained 
and may well be within the reach of the LHC and be produced copiously 
in pairs through its gluon interaction.

\noindent \underline{\it Two-Component Dark Matter}~:~
As shown in Ref.~\cite{dmt14}, the coexistence of the QCD axion with a 
stable weak-scale particle allows for a much more flexible two-component 
theory of dark matter.  It relaxes the severe constraints imposed on either 
component if considered alone.  It allows for a solution of the strong 
CP problem, without having the QCD axion as observable dark matter.
Regarding the weak-scale DM particle, it is the lightest particle charged under the residual dark parity $(-1)^{3B+L+2j}$, and can be either a real scalar \cite{Silveira:1985rk,McDonald:1993ex,Burgess:2000yq} or an admixture of a singlet-doublet fermion \cite{ArkaniHamed:2005yv,Mahbubani:2005pt,DEramo:2007anh,Enberg:2007rp}. 

In the scalar case, since $\chi_{1,2,3}$ carry lepton number, they are 
complex with invariant $(m_\chi^2)_{ij} \chi_i^* \chi_j$ terms. 
However, a $6 \times 6$ mass-squared matrix is obtained because there 
are also the allowed $\sigma^* \chi_i \chi_j$  terms.  There are thus 
six real scalar eigenstates.  
Since the mass splittings of the real and imaginary parts of the complex 
$\chi$ scalars are proportional to $\langle \sigma \rangle$, they 
are presumably large.    Hence fine tuning is 
required~\cite{dmt14} to make one component light and the other heavy, 
if we want the lightest $\chi$ (call it $\chi_0$) to be dark matter. 

In the fermion case, there are invariant mass terms $m_E \bar{E}_R E_L$ 
and $m_N N_L N_L$, as well as the allowed mixing terms between $N$ and 
$E^0$ which are proportional to $\langle \phi^0 \rangle$.  The $3 \times 3$ 
mass matrix spanning $(N_L, E^0_L, \bar{E}^0_R)$ is then of the form
\begin{equation}
{\cal M}_{NE} = \begin{pmatrix}m_N & m_L & m_R \\ m_L & 0 & m_E 
\\ m_R & m_E & 0\end{pmatrix}
\end{equation}
resulting in 3 Majorana fermion eigenstates, the lightest (call it $N_0$) 
is dark matter.  Assume for example $m_L=m_R=m_V/\sqrt{2} > 0$ and 
$m_N = m_E > 0$, then the three mass eigenvalues (in increasing magnitude) 
are
\begin{equation}
m_E - m_V, ~~~ -m_E, ~~~ m_E + m_V,
\end{equation}
corresponding to the three mass eigenstates
\begin{eqnarray}
N_0 &=& N_L/\sqrt{2} - (E^0_L + \bar{E}^0_R)/2, \\ 
N_1 &=& (E^0_L - \bar{E}^0_R)/\sqrt{2}, \\ 
N_2 &=& N_L/\sqrt{2} + (E^0_L + \bar{E}^0_R)/2. 
\end{eqnarray}

In either case, it may only account for part of dark matter, the rest 
coming from axions.  In direct-search experiments, the exchange of $Z$ 
is irrelevant because $\chi_0$ is a singlet, and $N_0$ is Majorana. 
However, the exchange of $h$ (the SM Higgs boson) will 
contribute.   As for relic abundance, beyond those interactions of 
the minimal models mentioned earlier, we have also the Yukawa terms 
$\bar{\nu}_i N_0 \chi$, which may also contribute.  There are many 
free parameters in our model to make this work, but it is not our 
goal to examine them in any detail.  After all, these issues have been 
dealt with thoroughly in those previous studies.  Instead, we will 
focus on the 
feasibility of finding the scalar quark $\zeta$ which connects the 
high scale ($10^9$ to $10^{11}$ GeV) of the axion to the much lower scale 
(100 GeV) of the dark-matter candidates $\chi_0$ and $N_0$.

\noindent \underline{\it Scotogenic Neutrino Mass}~:~
Using the Yukawa terms $\chi (\bar{\nu}_L E^0_R + \bar{e}_L E^-_R)$, 
$\bar{N}_L ( \phi^0 E^0_R - \phi^+ E^-_R)$, 
${N}_L ( \phi^0 E^0_L - \phi^+ E^-_L)$, and $\sigma_2^* \chi_i \chi_j$, 
\begin{figure}[htb]
\begin{center}
\includegraphics[scale=0.3]{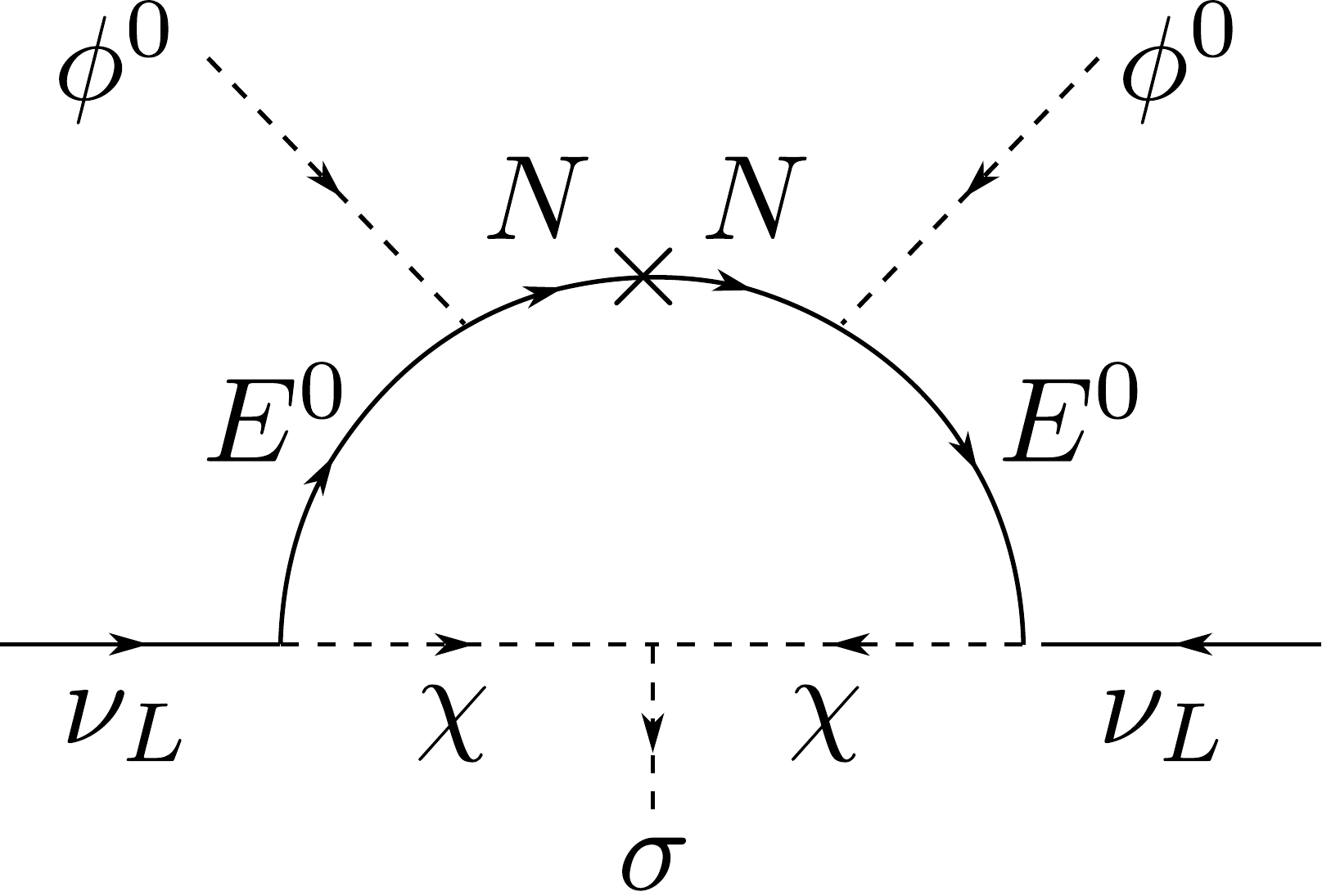}
\caption{One-loop generation of neutrino mass with $U(1)_L$.}
\end{center}
\end{figure}
the one-loop diagram of Fig.~1 is obtained, thereby generating three 
radiative neutrino masses through the spontaneous breaking of $U(1)_L$~\cite{fmp14,rrszt15}. 
This idea was previously applied directly to the canonical seesaw mechanism 
with singlet right-handed neutrinos~\cite{s87,m01},  
thus equating the 
axion scale to that of the neutrino seesaw.  Here the axion scale enters 
through $\langle \sigma \rangle$.  Our model differs conceptually from the 
previous use of Fig.~1 because we equate $U(1)_L$ with $U(1)_{PQ}$ and 
let it be spontaneously broken.  However, the resulting neutrino mass 
matrix has the same structure as previous studies and the details are 
available in those references~\cite{fmp14,rrszt15}.

In this scotogenic model, the family index is carried by $\chi$, so a 
possible family symmetry may be considered. 
Let $\nu_{1,2,3}$ and $\chi_{1,2,3}$ transform as $\underline{3}$ under 
the non-Abelian discrete symmetry $T_7$ for example~\cite{ckmo11}.  
The group multiplcation rule of $T_7$ is
\begin{equation}
\underline{3} \times \underline{3} = \underline{3} + \underline{3}^* 
+ \underline{3}^*, 
\end{equation}
and since $\sigma \sim \underline{1}$ under $T_7$, it does not close the 
loop of Fig.~1.  
We now need to add three extra scalars $\rho_{1,2,3} 
\sim \underline{3}$ or  $\underline{3}^*$ to couple to $\chi_i \chi_j$ to 
complete the loop of Fig.~1.  It is then possible for $\langle \rho_{1,2,3} 
\rangle$ to be much smaller than $\langle \sigma \rangle$, in which case 
the lightest $\chi$ may be naturally of the electroweak scale as a 
dark-matter candidate.

\noindent \underline{\it Possible Hadronic Connection}~:~
Whereas the heavy color-triplet fermion $D$ connects with the SM through 
the Yukawa term $\bar{D}_L d_R \chi$, another possible way is through the 
color-triplet scalar $\zeta$, with the important terms 
$f_D \zeta \bar{D}_L e_R + H.c.$ as well as
\begin{eqnarray}
& f_N \zeta \bar{u}_R N_L + f_E \zeta^* (\bar{E}^0_R u_L + \bar{E}^+_R d_L) 
+ H.c. = f_N \zeta \bar{u}_R (N_0 + N_2)/\sqrt{2}& \nonumber \\ 
& + f_E \zeta^*(-N_0/2 - 
N_1/\sqrt{2} + N_2/2) u_L + f_E \zeta \bar{d}_L E^-_R + H.c.&
\end{eqnarray}   
We assume that $D$ is very heavy, so it decays away quickly 
in the early Universe to either $e \zeta$ or $d \chi$.  Subsequently, 
either $\chi_0$ or $N_0$ becomes a component of the dark matter of the 
Universe, together with the axion.
  
To test our hypothesis, we propose a search for $\zeta$ at the LHC.
It is easily produced, because it is a scalar quark.  We assume first 
that the Majorana fermion $N_0$ is dark matter.  If $f_N$ is dominant, 
then $\zeta$ decays equally to $N_0$ and $N_2$, with a quark jet in 
each case.  Whereas $N_0$ is stable and invisible, $N_2$ will decay, i.e.
\begin{equation}
N_2 \to \nu_i \chi_j + E^\pm W^\mp
\end{equation}
with the subsequent decay or conversion (if $\chi_j$ is virtual)
\begin{equation}
\chi_j \to \nu_k N_0 + \ell_k^\pm E^\mp,
\end{equation}
and
\begin{equation}
E^\pm \to W^\pm N_0.
\end{equation}
Most events are then of the type 2 jets + missing energy.
They are thus analogous to scalar quark pair production with decays 
to a quark and a neutralino in supersymmetry.  
We can thus borrow from the existing studies of 
supersymmetric scenarios to put a bound on $m_\zeta$ as a function 
of $m_{N_0}$ and $m_{N_2}$.

If $f_E$ is dominant, then $\zeta$ decays equally to $N_{0,1,2}$ 
and $E^+$, with a quark jet in each case, as shown in Eq.~(7).  
Whereas most $N_{0,1,2}$ decays are invisible, $E^+$ decays according to 
Eq.~(10).  This is analogous to 
a squark decaying to a quark and a chargino which then decays to a $W$ 
and a neutralino in supersymmetry. If we 
focus on the leptonic decay of $W$, then the final states of 
$\zeta \zeta^*$ production at the LHC may also include 2 jets + 
$\ell^\pm$ + missing energy and 2 jets + $\ell^+_1 \ell^-_2$ + missing energy. 

Consider the alternative case that the real scalar $\chi_0$ is dark matter.  
This means that the fermions $N_i$ are heavier.  The decay of $\zeta$ to 
$u$ + $N_i$ will have another step, i.e. 
\begin{equation}
N_i \to \nu_j \chi_0,
\end{equation}
which are invisible.  The signature is 
again 2 jets + missing energy, but now there are two relevant masses, 
$m_{N_i}$ and $m_{\chi_0}$.  If $m_{N_i}$ is close to $m_\zeta$, the 
kinematics will be quite different from the supersymmetric analog 
discussed previously where $N_0$ is dark matter.  The quark jets 
will be soft and could miss the cut on their momenta.  In that case, if 
$f_N$ dominates in Eq.~(7), the signal is just missing energy.

If $f_E$ dominates, we have again the decay of $\zeta$ equally to $N_i$ and 
$E^+$.  For the latter, the second step is now
\begin{equation}
E^+ \to \chi_0 \ell_i^+.
\end{equation}
The final states of $\zeta \zeta^*$ production at the LHC will again 
include 2 jets + $\ell^\pm$ + missing energy and 
2 jets + $\ell^+_1 \ell^-_2$ + missing energy.  However, since the charged 
leptons come directly from $E$ decay, their numbers are not diminished 
by the branching fraction of $W$ to leptons as in the case where $N_0$ 
is dark matter.  Also, if $m_E$ is close to $m_\zeta$, the jets 
may be too soft to be observable.  In that case, we will only find leptons 
+ missing energy.

\noindent \underline{\it LHC Signatures}~:~
We will discuss first the case of fermion dark matter, where we have the main collider signature of two jets plus missing transverse energy: $2j+E_\textrm{T}^\textrm{miss}$. 
The further decay of $E^+$ to $E^0$ via a $W$-boson only increases the number of soft objects so that the main signal is still just  $2j+E_\textrm{T}^\textrm{miss}$. This contribution, through the corresponding quasi-degenerate $E^+$-$E^0$ states,   are usually already included in the analyses  to be discussed below.
 
\begin{figure}[t]
\begin{center}
\includegraphics[scale=0.55]{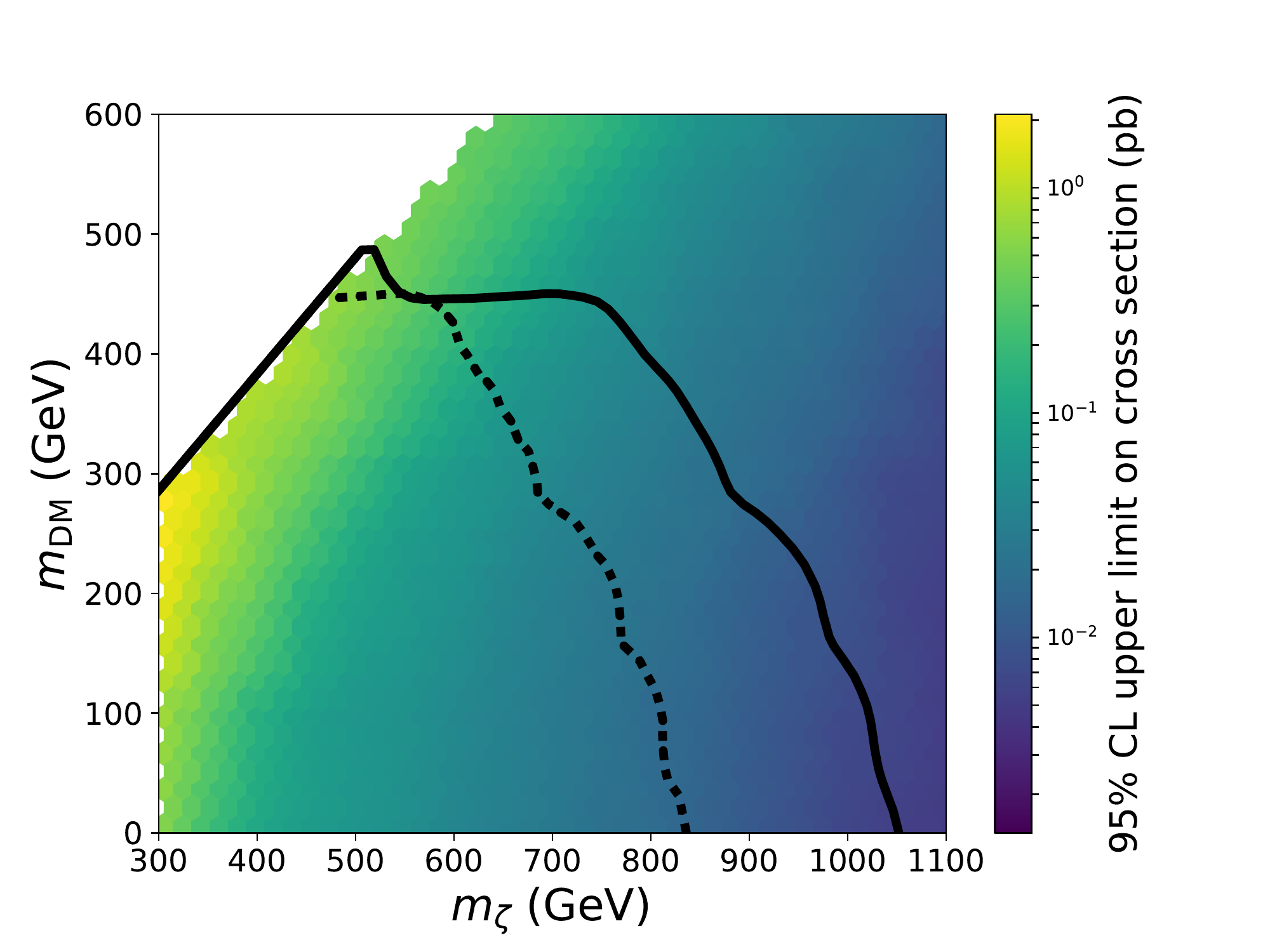}
\caption{Exclusion limit at 95\% CL on the cross section for $\zeta$ pair production from the CMS data \cite{CMS:2017kmd} based on 35.9 fb$^{-1}$ at 13 TeV. The region below the solid line corresponds to the excluded region for the case of fermion DM ($N_0$ is the DM particle) whereas the region below the dashed line is the excluded region for the case of scalar DM (when $\chi_0$ is the DM particle).}
\label{fig:exclusion8TeV}
\end{center}
\end{figure}

Regarding the signal   $2j+E_\textrm{T}^\textrm{miss}$, it has been already studied by ATLAS and CMS in the context of simplified supersymmetric scenarios searching for squarks decaying into a quark and neutralino.
In the case of fermion dark matter, our branching fraction is approximately $100\%.$  The results for the production of a single squark reported by 
CMS \cite{CMS:2017kmd} based on 35.9 fb$^{-1}$ at 13 TeV are thus fully applicable in our case and  reproduced in Fig.~\ref{fig:exclusion8TeV} (solid line).  They allow us to exclude, for example, $m_\zeta$ up to 1.0 (0.8) TeV for $m_{N_0}=$ 100 (400) GeV \footnote{It is worth noticing that contrary to the standard scenario of singlet-doublet fermion dark matter we can have now doublet-like fermion dark matter component less that 1~TeV while still being compatible with direct detection constraints.  
}.  
The exclusion limit at 95\% confidence level on the cross section of direct production of $\zeta$ pairs (color bar) from~\cite{CMS:2017kmd} is shown in Fig.~\ref{fig:exclusion8TeV}, where the region below the solid line corresponds to the current excluded region in the $(m_{N_0},m_\zeta)$ plane.  
At 13 TeV with 36.1 fb$^{-1}$ the ATLAS results \cite{ATLAS:2017cjl} are reported taking into account the production of 8 squark states of the first and second generation. Since the results are similar to the ones from CMS \cite{CMS:2017kmd}, we expect similar lower bounds on $m_\zeta$.  

In the case of scalar dark matter\footnote{We assume $m_N\gg m_E$.}, $E^0$
subsequently  decays to $\nu + \chi_0$
 and   $E^+$ to $\ell^+ + \chi_0$.  This leads to the following collider signatures: $2j+E_\textrm{T}^\textrm{miss}$, $2j+\ell^\pm+E_\textrm{T}^\textrm{miss}$ and  $2j+\ell_1^+\ell_2^-+E_\textrm{T}^\textrm{miss}$, with a branching fraction of $25\%$, $50\%$ and $25\%$, respectively.  
 These kinds of signals were studied in  Run-I by ATLAS at 8~TeV with luminosities  of around  $20~\text{fb}^{-1}$ in the context of simplified supersymmetric  models for squark production, assuming 100\% branching fractions  within each signal. 
 The three analysis are orthogonal and set independent limits on the squark and neutralino masses.
 We are not aware of published searches for first two-generation  squarks at Run-II in signals with hard-leptons, jets and  missing transverse momentum.
 ATLAS has searched for new phenomena in events of squark pair production having final states with  same-flavor opposite-sign dilepton pair, jets and large missing transverse momentum \cite{Aad:2015wqa,Aad:2015mia,Aad:2015iea}. The sequence of the decay processes is $\tilde{q}\tilde{q}^*\to(q\tilde{\chi}_2^0)(q\tilde{\chi}_2^0)$ with $\tilde{\chi}_2^0\to\tilde{\ell}^\mp\ell^\pm/\tilde{\nu}\nu$ and $\tilde{\ell}^\mp/\tilde{\nu}\to\ell^\mp\tilde{\chi}_1^0/\nu\tilde{\chi}_1^0$.

By comparing the results from ATLAS for   $2j+E_\textrm{T}^\textrm{miss}$ at 8~TeV \cite{Aad:2014wea} which uses a lepton veto, with the results of \cite{Aad:2015mia} which uses  additional leptons, we can check that the search without leptons has a greater sensitivity. For example,  for a neutralino mass of 100~GeV, the excluded mass of a eight-fold degenerate squark was 900~GeV without leptons, 860~GeV with opposite sign dileptons, and 800~GeV with one-lepton. In this way, the larger exclusion for the production of squarks  is in the signal without further leptons. In the case of scalar dark matter, the branching for two jets and zero leptons is $25\%$,
therefore the bounds discussed above at 13~TeV for this signal become weaker. In particular, we have found that the exclusion for $m_\zeta$ goes up to $\sim800$ ($600$) GeV for a DM mass of 100 (400) GeV. The full recast is presented in the lower exclusion curve (dashed line) of Fig. \ref{fig:exclusion8TeV}.

\begin{figure}[t]
\begin{center}
\includegraphics[scale=0.55]{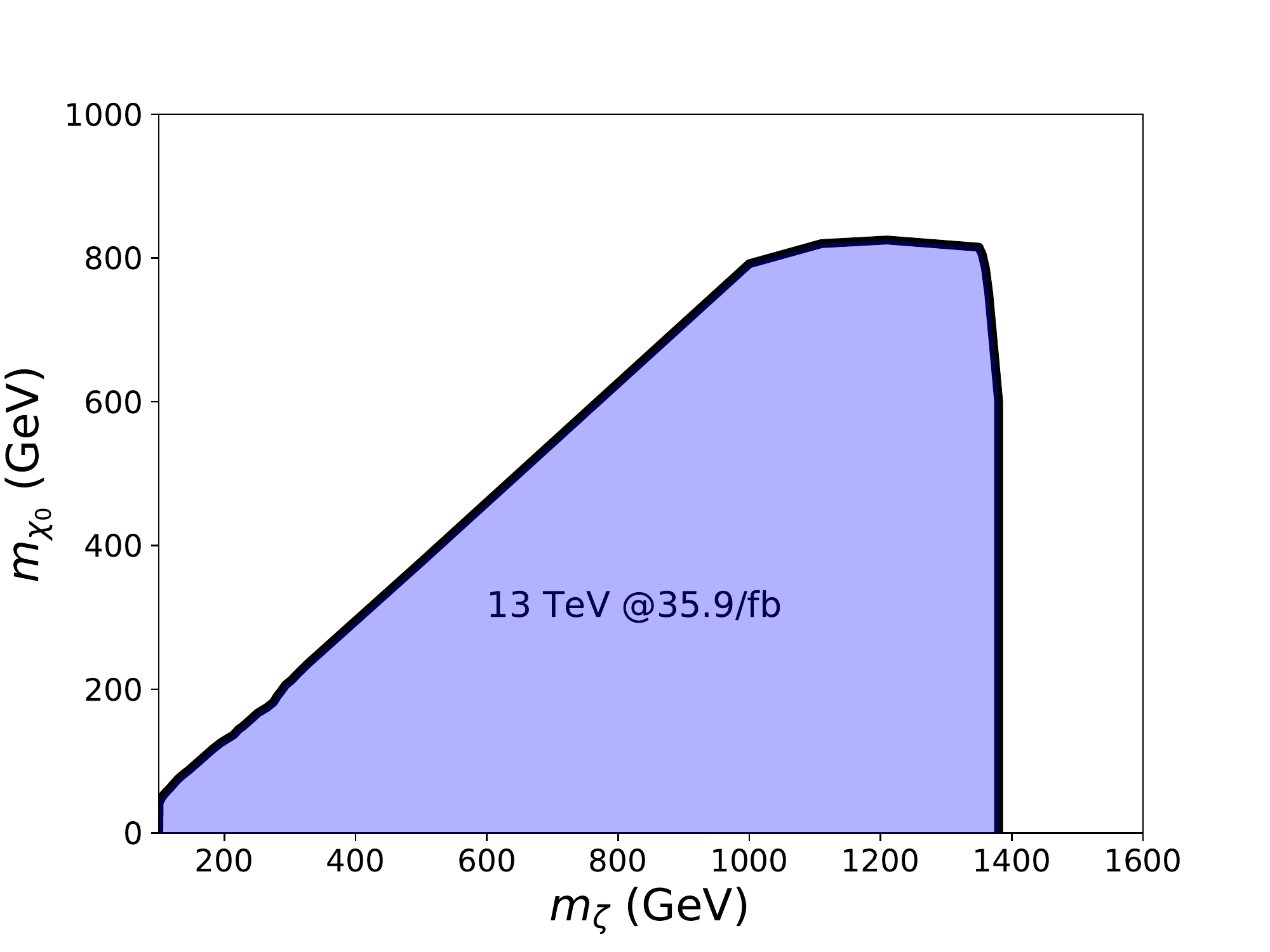}
\caption{Excluded region (below the solid line) for a singlet scalar quark decaying into a signal which becomes effectively of opposite-sign dileptons and missing transverse momentun. This is obtained from a recast of preliminary results of data for search of dileptons and missing transverse momentun  at 13 TeV by ATLAS~\cite{ATLAS:2017uun} by using a luminosity of $35.9\ \text{fb}^{-1}$.} 
\label{fig:exclusion}
\end{center}
\end{figure}

In our model the greater exclusion happens when  the mass of $E^+$ is close to the $\zeta$ mass, such that the jets are sufficiently soft, so that the signal becomes effectively   $\ell^+\ell^-+E_\textrm{T}^\textrm{miss}$ without jets.
In such a case, we can recast the searches for simplified supersymmetric models with slepton pair production. 
The results of searches for the first two generations of sleptons at Run-II in signals with opposite sign dileptons and  missing transverse momentum for $36.1\ \text{fb}^{-1}$ are reported by ATLAS in \cite{ATLAS:2017uun}.
Taking into account the $25\%$  branching into  the charged lepton and the scalar dark matter particle, the exclusion region at 13 TeV  covers mass values up to $m_\zeta\sim1400$ GeV for $m_{\chi_0}\lesssim800$~GeV. 
In  Fig.~\ref{fig:exclusion}, we present the full recasted exclusion region at 13~TeV. 

The limits with one additional lepton studied in \cite{Aad:2015mia} are applicable to the case where $N_L$ is the lightest or the next to lightest neutral fermion, since this would correspond to an intermediate $\tilde{\chi}_2^0$ which decays into $\chi_1^0$  with further gauge bosons.  Therefore, we would expect softer bounds in this case. 

\noindent \underline{\it Conclusion}~:~
A new insight as to the nature of lepton number has been proposed.  It is identified with the Peccei-Quinn symmetry which solves the strong CP problem, with the appearance of an invisible axion.  A residual symmetry remains, i.e. $(-1)^L$, which serves also as dark parity, i.e. $(-1)^{3B+L+2j}$.  New particles which are odd under this $Z_2$ allow the one-loop radiative generation of neutrino masses, and provide a weak-scale component of dark matter in addition to the axion. We show how the two sectors may be connected with a new singlet scalar quark $\zeta$, which may be easily probed (or discovered in the future) at the LHC through its subsequent decays to either the fermion or scalar dark-matter candidate. 

\noindent \underline{\it Acknowledgements}~:~
The work of E. M. has been supported in part by the U.~S.~Department of Energy under Grant No.~DE-SC0008541. 
The work of D. R. and O. Z. has been partly supported by the Grants Sostenibilidad-GFIF and CODI-IN650CE, and by COLCIENCIAS through the Grant No. 111-565-84269.

\bibliographystyle{unsrt}

\end{document}